\documentstyle[preprint,aps]{revtex}

\preprint{Int. J. of Bif. and Chaos}

\begin{document}


\title{Transition to Chaotic Phase Synchronization through \\
Random Phase Jumps}

\author{D. Paz\'o\footnote{E-mail: diego@fmmeteo.usc.es; http://fmmeteo.usc.es}, I.P. Mari\~no, 
V. P\'erez-Villar and V.\ P\'erez-Mu\~nuzuri}
\address{Group of Nonlinear Physics, Faculty of Physics, \\
University of Santiago de Compostela, \\
15706 Santiago de Compostela, Spain}

\date{\today}

\maketitle

\begin{abstract}
Phase synchronization is shown to occur between opposite cells of a ring consisting of chaotic Lorenz
oscillators coupled unidirectionally through driving. As the coupling strength is diminished, full
phase synchronization cannot be achieved due to random generation of phase jumps. The brownian dynamics
underlying this process is studied in terms of a stochastic diffusion model of a particle in a 
one-dimensional medium.
\end{abstract}


\newpage

\section{Introduction}

Phase synchronization phenomena in coupled chaotic systems have been extensively studied during
the last few years in the context of non-identical chaotic systems (Rosenblum et al., 1996; Osipov et al., 1997), 
ecological systems (Earn et al., 1998; Blasius et al., 1999), physiological systems (Sch\"afer et al., 1998), 
chaotic systems forced by an external periodic or noisy signal (Pikovsky et al., 1997a, 1997b), 
an ensemble of coupled chaotic oscillators (Pikovsky et al., 1997c; Osipov et al., 1997),
and with an electronic model of two R\"ossler oscillators (Parlitz et al., 1996).
This effect owes its name from the classical definition of synchronization of periodic oscillators
which is described in terms of locking or entrainment of the phases, while the amplitudes can
be quite different. Hence, synchronization of chaotic oscillators can be defined in the most general case,
as the locking between the phases of two coupled systems, while the amplitudes remain chaotically 
varying in time (Rosenblum et al., 1996).

For chaotic oscillators, there is no unique definition of phase. An approach to 
determine the amplitude $A$ and phase $\phi$ of a narrow-band signal $s(t)$ is based on the analytic signal concept that
considers an analytical signal $\psi(t)$ as a complex function of time, $\psi(t)=s(t) + \imath \tilde{s}(t) =
A(t) e^{\imath \phi(t)}$ and $\tilde{s}(t)$ is the Hilbert transform of $s(t)$ (Rosenblum et al., 1996). 
However, in other cases phase and
amplitude can be defined as a function of the natural variables of the oscillator. For example, for the R\"ossler
attractor $\phi=\arctan(y/x)$ (Pikovsky et al., 1997d) or 
$\phi=\arctan(y/\sqrt{x^2+y^2})$ (Rosa et al., 1998), and for the Lorenz model
$\phi=\arctan\left[(\sqrt{x^2+y^2} - u_0)/(z-z_0)\right]$ (Pikovsky et al., 1997b), where $u_0$ and $z_0$ are constants.

In this paper, we focus our interest in the phenomenon of phase synchronization between 
chaotic Lorenz systems coupled unidirectionally through driving in a ring geometry. It has been shown, that for
an appropriate set of parameters, a ring of $N$ coupled Lorenz systems shows a Periodic Rotating Wave (PRW) where
neighboring oscillators exhibit a phase difference of $2\pi/N$ and the amplitude varies with time sinusoidally
(Mari\~no et al., 1998). This system, with a different set of parameters also
exhibits Chaotic Rotating Waves (CRW) defined as well by a phase difference between neighboring cells of $2\pi/N$ but
the amplitude remains chaotic (S\'anchez and Mat\'{\i}as, 1999). In this structure there exists a superposition of Fourier 
modes $k=0$ and $k=1$. Here we will show a transition from a PRW with a phase difference of $2\pi/N$ to
a CRW with a phase difference of $4\pi/N$, where opposite cells are phase synchronized ($N$ even). Depending on the 
unidirectional coupling strength, random/brownian $2\pi$-phase slips develop during the mentioned transition.
\section{Model}

We shall consider rings of Lorenz attractors coupled in such a way that the dynamical
behavior (G\"u\'emez and Mat\'{\i}as, 1995) is defined by,
\begin{eqnarray}
\dot{x_j}&=&\sigma(y_j-x_j) \nonumber\\
\dot{y_j}&=&R\,\left(\beta\,x_{j-1}+(1-\beta)\,x_j\right)-y_j-x_j\,z_j \label{eqlor}\\
\dot{z_j}&=&x_j\,y_j-b\,\,z_j \nonumber
\end{eqnarray}
with $\sigma$, $R$ and $b$ positive parameters. Usual parameters values
are $\sigma=10$, $b=\frac{8}{3}$ and $R=28$. In Eq.~(\ref{eqlor}), $\beta$ accounts for the coupling strength,
$j$ runs from 1 to $N$ (number of cells in the array), and for $j=1$, $x_0=x_N$.

For $\beta=1$, it was observed (Mat\'{\i}as et al., 1997; Mari\~no et al., 1998) 
that the synchronized chaotic state is stable
if the size of the ring is small enough $N=2$, while for a certain critical number $N_c=3$ in the case of the Lorenz
model, an instability associated to the first Fourier mode $k=1$ 
destroys the uniform chaotic state, leading to a PRW. As the size of the ring is increased, new Fourier modes
become unstable and for $N=6$ a second instability ($k=2$) develops that could lead to a Chaotic Rotating Wave (CRW) where
neighboring oscillators exhibit a phase difference of $4\pi/N$ as it is shown in Fig.~\ref{fig_phase}(a), that is, 
Fourier modes $k=1$ and $k=2$ compete in a nonlinear way.  
Thus, opposite cells are phase synchronized while amplitudes remain chaotic and are, in general, uncorrelated. 
Figure~\ref{fig_phase}(b) shows the uncorrelated values of the amplitudes of 
$x_{j+N/2}(t)$ as a function of $x_{j}(t)$. Since the second Fourier mode plays an important
role for phase synchronization, we will focus our study in a ring consisiting of $N=6$ Lorenz cells described
by Eq.~(\ref{eqlor}) as $\beta$ is varied. This phase synchronization describes the onset of long-range
correlations in chaotic oscillations (suppression of phase diffusion), and thus also corresponds to the appearance of certain
order inside chaos that here is shown as a CRW with certain similarities to a quasiperiodic motion.

To study phase synchronization of coupled chaotic systems, we calculate the phases of the oscillators and
then check whether the weak locking condition $\Delta\phi=\vert n\phi_j-m\phi_{j+N/2}\vert <\,$const is satisfied. 
In this paper, we restrict
ourselves to the case of $m=n=1$. 
The definition of the phase for a given oscillator may be a problematic task as soon as there is not a center of rotation. 
Fig.~\ref{fig_lissajous} shows the $(x,y)$ projection of an oscillator phase space for
two $\beta$ values, $\beta=1.0$ and $\beta=0.85$. 
At $\beta=1.0$ a center of rotation can be clearly distinguished at $(x,y)=(0,0)$.
Then a Poincar\'e surface of section $y=x$, $x>0$ allows us to define the phase as (Pikovsky et al., 1997b) 
\begin{equation}
\phi(t) = n + { {t-t_n}\over{t_{n+1}-t_n}}
\label{eq_phase}
\end{equation} 
where $t_n$ is the nth crossing of the surface. Note that the phase has beeen normalized by a factor $2\pi$. We see
 that
with the surface chosen in Fig. 2(a) crossings are equal to maxima of the variable $x(t)$. Therefore, we can know 
at
what times the phase is an integer just looking at the time evolution of the variable $x$, this criterion has
been used before (Blasius et al., 1999).
As long as $\beta$ decreases
unproper rotations become more frequent, (see Fig.~\ref{fig_lissajous}(b)). However it is clear the existence 
of a "rotation axis", unlike the funnel R\"ossler attractor case (see e.g. 
Pikovsky et al., 1997b) where an independent center of rotations emerges at one side of the attractor. In consequence
phase is increased in one unit (i.e. $2\pi$ radians) when an unproper rotation occurs. 
As it will be later discussed the time difference of two consecutive maxima does not depend on the own nature of each rotation.
So it seems that our definition provides a "good" period. 
The instantaneous phase $\phi(t)$ will be determined through linear interpolation after calculating the 
instants of time at which maxima appear in the $x(t)$ series. 

\section{Results}

The main effect of varying $\beta$ is shown in Fig.~\ref{fig_temporal}. As shown above, for $\beta=1$ opposite oscillators
within the ring are phase synchronized. As $\beta$ decreases, opposite cells still remain phase synchronized, except for some phase
jumps. These events are defined as the non-occurrence of a maxima at its due time
in one of the $x(t)$ signals corresponding to cells $j$ or $j+N/2$. In other words, we will assume that a phase 
jump occurs if $\Delta\phi(t)=\phi_j(t)-\phi_{j+N/2}(t)$ changes by approximately $\pm 1$ 
for two consecutive maxima of $x(t)$.
Note for example for $\beta=0.76$ and $\beta=0.85$ two jumps are encircled in Fig.~\ref{fig_temporal}.
Unproper rotations are signaled, it is clear that they do not produce phase slips, although both phenomena appear
when the signal turns more chaotic.
Further decreasing the coupling strength $\beta$
finally leads to the formation of a PRW with a phase difference of $1/6$ (mod 1) between neighboring cells. The transition
between PRWs and phase synchronization with {\em jumps} occurs for a critical value of $\beta_c \approx 0.75$. 

The distribution of phase jumps is shown in the sequence of 
figures at the rightside of Fig.~\ref{fig_temporal} where the periods of time $T$ between consecutive maxima of $x(t)$ are
shown for two opposite cells within the ring. As $\beta$ is decreased, the map $T_{j+N/2}$ as a function of $T_{j}$ shows a greater 
dispersion from the mean value (located in the center of the figures) until the critical value $\beta_c$ is reached.
For high values of $\beta$, a small deviation of periods around the mean value appears according to the way the 
phase synchronization has been defined ($\vert\phi_1 - \phi_4\vert <\,$ const $< 1/2$). As the value of $\beta$ decreases, the
dispersion around the mean value increases at the same time that two independent accumulation regions responsible for the
phase slips appear (see circles at the rightside of Fig.~\ref{fig_temporal}). 
Notice that these phase slips are not related to the unproper rotations which are represented by maxima (minima) peaks
of the temporal serie that do not take a positive (negative) value (see arrows in the leftside of Fig.~\ref{fig_temporal}). 
The process of phase jumps formation
is as follows; consecutive $x(t)$ maxima of one cell remains phase synchronized with the opposite cell within the ring,
until a phase jump occurs spontaneously, which corresponds to jumps from arms numbered (2) and (3) in Fig.~\ref{fig_temporal}
to the encircled zones. That is, phase slips are characterized by the sequences $2 \rightarrow 1 \rightarrow 2$ and
$3 \rightarrow 4 \rightarrow 3$. At the same time, fluctuations in $\Delta\phi$ (i.e. no perfect synchronization between maxima)
leads to jumps between zones (2) and (3). Besides, it must be pointed out that as $\beta$ is
decreased the concept of phase synchronization defined above and used here becomes less restrictive as the dispersion around
the mean value increases (const $\rightarrow 1/2$). 

The number of phase slips occurring at a given interval of time decreases as $\beta$ is increased. 
Then, when opposite cells within the ring are phase synchronized the phase difference $\vert\Delta\phi(t)\vert$ is, on average,
constant in time. But, if phase slips occur for $\beta_c < \beta < 1$, then one would expect that for $M$ different 
initial conditions, the averaged square phase difference dynamics will be generally diffusive, so for large $t$,
\begin{equation}
\langle \vert\Delta\phi(t)\vert^2 \rangle = 2\,D\,t
\label{eq_difusion}
\end{equation}
where $D$ is the diffusion constant. Figure~\ref{fig_promedio} shows the linear dependence found for the
root mean square of the phase difference $\langle \vert\Delta\phi(t)\vert^2 \rangle^{1/2}$ 
as a function of time, in a log-log plot for $M=14$ different random initial conditions for 
the Lorenz cells within the ring and for four different values of $\beta$. The four graphs fit to a straight line
with slope $S \approx 1/2$ as expected from Eq.~(\ref{eq_difusion}) (see Table I for the fitted values). 

The distribution of temporal periods $\tau$ between two consecutive phase jumps is shown in Fig.~\ref{fig_tau} 
for two different values of $\beta$. Note the occurrence of longer periods of time $\tau$ for higher values of $\beta$.
These distributions show an exponential decay with $\tau$ as a consequence of the intrinsic random/brownian nature of the
dynamical process underlying the formation of phase slips. Moreover, neither 
the phase jumps occur simultaneously for all couples of opposite cells within the array, nor the phase jumps
are correlated in space, which is in agreement with the random dynamics of phase slip formation. 

\section{Discussion}

From Fig.~\ref{fig_tau} a mean value of the period $\langle\tau\rangle$ for each value of $\beta$ can be defined. 
Now, by using a simple model of stochastic diffusion of a particle in a one-dimensional
medium ({\em random discrete walk}), the averaged quadratic dispersion from the phase synchronized state ($\Delta\phi \approx 0$)
is given by the following equation,
\begin{equation}
\langle \vert\Delta\phi(t)\vert^2 \rangle = \frac{t}{\langle\tau\rangle}
\label{eq_walk}
\end{equation}
where $t/\langle\tau\rangle$ is the number of phase jumps that have appeared for $t \gg \langle\tau\rangle$. 
Consequently, comparing Eqs.~(\ref{eq_difusion}) and (\ref{eq_walk}), it is possible to calculate a theoretical value for 
the diffusion coefficient $D_{th}=(2\langle\tau\rangle)^{-1}$. A comparison between the diffusion coefficient $D_{exp}$ obtained
after fitting the log-log plots given in Fig.~\ref{fig_promedio} and $D_{th}$ is shown in Table I. 
Note the good agreement between both 
coefficients for large values of $\beta$ as expected for a typical brownian dynamics.
It must be noted that $\langle\tau\rangle$ increases dramatically with $\beta$ (see Fig.~\ref{fig_tau} 
and the values of $D_{th}$ in Table I) in such a way that it is not possible to assure the existence
of an upper limit of $\beta$ above it no jumps appear.
For $\beta \rightarrow \beta_c$ we have found small values of $\langle\tau\rangle$ 
of the order of the mean period between two consecutive maxima
of $x(t)$. Thus, jumps occur frequently in time and a random, uncorrelated in time, sequence cannot be assured
(the system shows a tendence to display $+1,\,-1,\,+1,\ldots$ slips series). Then,
the obtained values of the diffusion $D_{exp}$ are smaller than those predicted $D_{th}$ using $\langle\tau\rangle$.

The transition between periodic rotating waves and phase synchronized chaotic rotating waves has been shown to occur
as the coupling strength $\beta$ is increased. For values of $\beta>\beta_c$, phase slips develop randomly in time 
following a diffusive process given by Eq.~(\ref{eq_difusion}). Note that the dynamics of the phase defined for a single
chaotic oscillator is generally diffusive as well (Pikovsky et al., 1997b) and in this case, $D$ determines the phase
coherence of the chaotic oscillations which is inversely proportional to the width of the spectral peak
of the chaotic attractor. On the other hand, for coupled unsynchronized nonidentical 
chaotic oscillators the average phase difference
grows linearly with time (Blasius et al., 1999). Nevertheless, we have shown a different behavior where the 
root mean square of the phase difference grows with $t^{1/2}$ as a consequence of phase slips random formation.

\section*{Acknowledgements} 

We want to thank I. Sendi\~na-Nadal for fruitful discussions and comments on this work.
The support by DGES and Xunta de Galicia under Research Grants 
PB97--0540 and XUGA--20602B97, respectively, is gratefully acknowledged.

\section*{References}
\begin{description}

\item Blasius, B., Huppert, A. and Stone L. [1999] "Complex dynamics and phase synchronization in spatially 
extended ecological systems". {\it Nature} {\bf 399}, 354-359.

\item Earn, D.J. D., Rohani, P. and Grenfell, B. [1998] "Persistence, chaos, and synchrony in ecology and epidemiology".
{\it Proc. R. Soc. Lond. B} {\bf 265}, 7-10.

\item G\"u\'emez, J. and Mat\'{\i}as, M.A. [1995] "Modified method for synchronizing and cascading chaotic systems".
{\it Phys. Rev. E} {\bf 52}, 2145-2148.

\item Mari\~no, I.P., P\'erez-Mu\~nuzuri, V. and Mat\'{\i}as, M.A. [1998] "Desynchronization transitions in rings 
of coupled chaotic oscillators". {\it Int. J. of Bif. and Chaos} {\bf 8}, 1733-1738.

\item Mat\'{\i}as, M.A., P\'erez-Mu\~nuzuri, V. Lorenzo, M.N., Mari\~no, I.P. and P\'erez-Villar, V. [1997] 
"Observation of a fast rotating wave in rings of coupled chaotic oscillators". {\it Phys. Rev. Lett.} {\bf 78}, 219-222.

\item Osipov, G.V., Pikovsky, A.S., Rosenblum, M.G. and Kurths J. [1997] "Phase synchronization effects in a lattice of 
nonidentical R\"{o}ssler oscillators". {\it Phys. Rev E} {\bf 55}, 2353-2361.

\item Parlitz, U., Junge, L., Lauterborn, W. and Kocarev, L. [1996] "Experimental observation of phase
synchronization". {\it Phys. Rev. E} {\bf 54}, 2115-2117.

\item Pikovsky, A., Osipov, G., Rosenblum, M., Zaks, M. and Kurths, J. [1997a] "Attractor-repeller collision and eyelet 
intermittency at the transition to phase synchronization". {\it Phys. Rev. Lett.} {\bf 79}, 47-50.  

\item Pikovsky, A., Rosenblum, M., Osipov, G., and Kurths, J. [1997b] "Phase synchronization of chaotic oscillators 
by external driving". {\it Physica D} {\bf 104}, 219-238.

\item Pikovsky, A., Rosenblum, M.G. and Kurths, J. [1997c] "Synchronization in a population of globally coupled chaotic 
oscillators". {\it Europhys. Lett.} {\bf 34}, 165-170. 

\item Pikovsky, A., Zaks, M., Rosenblum, M., Osipov, G. and Kurths, J. [1997d] "Phase synchronization of chaotic 
oscillations in terms of periodic orbits". {\it Chaos} {\bf 7}, 680-687.

\item Rosa, E., Ott, E. and Hess, M.H. [1998] "Transition to phase synchronization of chaos". {\it Phys. Rev. Lett.}
{\bf 80}, 1642-1645.

\item Rosenblum, M.G., Pikovsky, A.S. and Kurths, J. [1996] "Phase synchronization of chaotic oscillators". 
{\it Phys. Rev. Lett.} {\bf 76}, 1804-1807.

\item S\'anchez, E., Mat\'{\i}as, M.A. [1999] "Transition to rotating chaotic waves in arrays of coupled Lorenz oscillators".
{\it Int. J. of Bif. and Chaos} {\bf 9} (in press).

\item Sch\"afer, C., Rosenblum, G.R., Kurths, J. and Abel, H.H. [1998] "Heartbeat synchronized with ventilation". 
{\it Nature} {\bf 392}, 239-240.

\end{description}


\begin{figure}
\caption[]{Temporal evolution of the variable $x$ of three contiguous Lorenz cells coupled unidirectionally through
driving within a ring of $N=6$ oscillators and $\beta=1$ (a). The figure shows a Chaotic Rotating Wave (CRW) with approximate
phase relationship of $4\pi/N$. Variables $x_j$ and $x_{j+N/2}$ ($j=1$) are then phase correlated although their amplitudes
remain chaotic, and uncorrelated (b).}
\label{fig_phase}
\end{figure}

\begin{figure}
\caption[]{Projections onto the $x-y$ plane of the trajectory followed by a chaotic oscillator when considering an
array of six identical cells coupled with different values of the coupling parameter: 
(a) $\beta=1.0$ and (b) $\beta=0.85$. In  (a) the straight line indicates the Poincar\'e surface 
of section $y=x$, $x>0$, while in (b) it represents the "rotation axis", since now a rotation center
is not properly defined.}
\label{fig_lissajous}
\end{figure}

\begin{figure}
\caption[]{Temporal evolution of the variables $x_1$ and $x_4$, corresponding to two opposite cells within
a ring of $N=6$ Lorenz oscillators, (leftside) and temporal period $T_4$ between maxima of $x_{4}(t)$ as a function of $T_1$
(rightside) for four different values of $\beta$. Note in the leftside of the figure that the phase slips are encircled.
For $\beta=0.90$ the arrows indicate the time intervals where the phase is not properly defined (here the
local minima are  positive). Note that these positive  minima   are not related to  the encircled phase slips.

At the rightside, circles mark those accumulation regions where the periods are greater than the average value, signalling
then the occurrence of a phase jump. See text for an explanation of the sequence of numbers: $1 \rightarrow 4$. 
For $\beta=0.75$ the arrow indicates the position of a single dot corresponding to
the value of $T_1=T_4=0.36$ t.u.}
\label{fig_temporal}
\end{figure}

\begin{figure}
\caption[]{Log--log plot of the root mean square of the phase difference $\langle\vert\Delta\phi\vert^{2}\rangle^{1/2}$ 
as a function of time for four different values of the coupling strength $\beta$. Lines correspond to a linear fitting.}
\label{fig_promedio}
\end{figure}

\begin{figure}
\caption[]{Distribution of temporal periods $\tau$ between two phase jumps for two different values of $\beta$.
Note the different time scales between both graphs. Fitting lines correspond to an exponential decay (first order in time).}
\label{fig_tau}
\end{figure}



\newpage

\begin{center}
\begin{tabular}{|c|c|c|c|}
\hline
\hline
$\beta$ & $S$ & $D_{exp} \times 10^{-3}$ (t.u.$^{-1}$) & $D_{th} \times 10^{-3}$ (t.u.$^{-1}$) \\
\hline
0.85\, & $0.51 \pm 0.02$\, & $3.53 \pm 2.35$ & $4.41 \pm 0.05$ \\
0.86\, & $0.47 \pm 0.01$\, & $2.98 \pm 1.09$ & $2.32 \pm 0.03$ \\
0.87\, & $0.48 \pm 0.02$\, & $1.55 \pm 1.02$ & $0.73 \pm 0.02$ \\
0.88\, & $0.52 \pm 0.01$\, & $0.050 \pm 0.018$ & $0.055 \pm 0.001$ \\
0.89\, & $0.48 \pm 0.01$\, & $0.006 \pm 0.002$ & $0.004 \pm 0.001$ \\
\hline
\hline
\end{tabular} \\
\vspace{.5 cm}
TABLE I: Values of the slope $S$, and $D_{exp}$ obtained from the linear fitting of Fig.~\ref{fig_promedio} and
Eq.~(\ref{eq_difusion}). $D_{th}=(2\langle\tau\rangle)^{-1}$ is calculated from the mean values of $\tau$.
\end{center}

\end{document}